\documentclass{jpsj-suppl}
\usepackage{txfonts}

\title{Hysteresis and Relaxation Effects in the Spin-Ice Compound $\text{Dy}{}_\text{2}\text{Ti}_\text{2}\text{O}_\text{7}$ studied by Heat Transport}

\author{Simon \textsc{Scharffe}, Gerhard \textsc{Kolland}\thanks{Present address: Deutsches Zentrum f\"{u}r Luft- und Raumfahrt, Linder H\"{o}he,  51147 K\"{o}ln, Germany}, Martin \textsc{Hiertz}, Martin \textsc{Valldor}\thanks{Present address: Max-Planck-Institut f\"{u}r Chemische Physik fester Stoffe, Noethnitzer Str. 40, 01187 Dresden, Germany}, and Thomas \textsc{Lorenz}
}

\inst{II. Physikalisches Institut, Universit\"{a}t zu K\"{o}ln, Z\"{u}lpicher Str. 77, 50937 K\"{o}ln, Germany}

\email{scharffe@ph2.uni-koeln.de}

\recdate{September 30, 2013}

\abst{The low-temperature thermal conductivity $\kappa$ of the spin-ice compound $\text{Dy}{}_\text{2}\text{Ti}_\text{2}\text{O}_\text{7}$ shows pronounced hysteresis as a function of magnetic field. Here, we investigate how these hysteresis effects depend on temperature, the magnetic-field direction, the rate of magnetic-field change, and on the direction of the heat current. In addition, the time-dependent relaxation of the heat conductivity is investigated. These measurements yield information about possible equilibrium states and reveal that in the low-field and low-temperature region extremely slow relaxation processes occur. 
}

\kword{spin ice, $\text{Dy}{}_\text{2}\text{Ti}_\text{2}\text{O}_\text{7}$, magnetic frustration, relaxation, magnetic heat transport}

\begin{document}
\maketitle

\section{Introduction}

The study of magnetic monopole excitations in the spin-ice compound $\text{Dy}{}_\text{2}\text{Ti}_\text{2}\text{O}_\text{7}$ is of high interest in current solid-state research~\cite{Ryzhkin2005,Castelnovo2008,Morris2009,Giblin2011,Castelnovo2011,Kadowaki2009,Jaubert2011,Bramwell2009,Blundell2012,Yaraskavitch2012}. $\text{Dy}{}_\text{2}\text{Ti}_\text{2}\text{O}_\text{7}$ is a geometrically frustrated spin system where the magnetic $\text{Dy}^\text{3+}$ ions form a pyrochlore lattice of corner-sharing tetrahedra. A strong crystal field results in an Ising anisotropy with the local easy axes along one of the \{111\} directions. As a consequence, the $\text{Dy}^\text{3+}$ moments point either into or out of each tetrahedron and allow several degenerate spin configurations. When two spins point in and two out of a tetrahedron the magnetic dipole energy is minimized, leading to a sixfold degenerate ground state in zero magnetic field~\cite{Ramirez1999,Bramwell2001,Sakakibara2003,Nagle1966,Meng2013}. This orientation of the $\text{Dy}^\text{3+}$ moments corresponds to the hydrogen displacement of water ice. Excited states are created by flipping a single spin resulting in two neighboring tetrahedra with configurations \mbox{1in-3out} and \mbox{3in-1out}, respectively. Due to the ground-state degeneracy such a dipole excitation fractionalizes into two individual monopole excitations, which can freely propagate in zero field.

Recently, we investigated these exotic magnetic excitations by thermal-conductivity measurements and found experimental evidence for monopole heat transport in $\text{Dy}{}_\text{2}\text{Ti}_\text{2}\text{O}_\text{7}$ \cite{Kolland2012,Kolland2013}. In zero magnetic field, this leads to a pronounced magnetic contribution \mbox{$\kappa_\text{mag}$} in the low-temperature thermal conductivity $\kappa$. For finite magnetic fields, \mbox{$\kappa_\text{mag}$} decreases and we observed a correlation between the magnitude of \mbox{$\kappa_\text{mag}$} and the degree of degeneracy of the different magnetic-field induced ground states for \mbox{$\vec H \,||\, [001]$},  \mbox{$\vec H \,||\, [1\bar{1}0]$}, and  \mbox{$\vec H \,||\, [111]$}. For all three field directions pronounced hysteresis effects occur between the $\kappa(H)$ curves measured with increasing or decreasing magnetic field. Below about $0.6 \, \text{K}$, we observed that as a function of decreasing field $\kappa(H \rightarrow 0)$ does not recover the initial \mbox{$\kappa_0=\kappa(H=0)$} obtained by zero-field cooling. Moreover, we found that the reduced \mbox{$\kappa(H \rightarrow 0)$} values slowly relax towards the corresponding $\kappa_0$. Slow relaxation phenomena in the low-temperature region have also been observed in other physical properties of $\text{Dy}{}_\text{2}\text{Ti}_\text{2}\text{O}_\text{7}$ , \emph{e.g.}, in the magnetization \cite{Matsuhira2011}, the a.c.-susceptibility \cite{Yaraskavitch2012} or the specific heat \cite{Klemke2011,Kolland2012,Meng2013}.

An additional hysteresis of $\kappa(H)$ is observed in the so-called kagome-ice state, which is realized at low temperature for  \mbox{$\vec H \,||\, [111]$} below about \mbox{$1 \, \text{T}$}. For this field direction, the pyrochlore structure can be best visualized as alternating triangular and kagome planes of Dy spins, which are stacked along $[111]$. While the spins of the triangular planes are fully aligned already by small fields  \mbox{$\vec H \,||\, [111]$}, the competition between the 2in/2out ice rule and the Zeeman energy results in a threefold degenerate kagome-ice state, where in every tetrahedron 1 out of the 3 Dy moments of the kagome plane has a finite component opposite to the magnetic-field direction. At low temperature, the kagome-ice phase is characterized by a plateau in the magnetization $M(H)$ up to about $1 \, \text{T}$, where a transition to the fully polarized state occurs. Surprisingly, $\kappa(H)$ is strongly hysteretic within the kagome-ice phase~\cite{Kolland2013,Sun2013}, whereas there is no hysteresis in the corresponding magnetization plateau. This unusual hysteresis of $\kappa(H)$ has been observed in Ref.~\cite{Sun2013} in measurements with the heat current $\vec{\jmath}$ along and perpendicular to the magnetic field, \emph{i.e.} with $\vec{\jmath}$ perpendicular and within the kagome planes, respectively, and has been confirmed by our data~\cite{Kolland2013} measured with \mbox{$\vec \jmath \,||\, [1\bar{1}0]$}. However, according to Ref.~\cite{Sun2013} no hysteresis  of the zero-field values $\kappa(H=0)$ seems to be present for both directions of $\vec{\jmath}$, in contrast to our results of Ref.~\cite{Kolland2013} for \mbox{$\vec \jmath \,||\, [1\bar{1}0]$}.

In this report, we present a study of these unusual hysteresis effects of  $\kappa(H)$ for the magnetic field directions along $[001]$ and $[111]$. In order to derive in how far the different $\kappa$ values represent equilibrium values, we studied $\kappa(H)$ for different magnetic-field sweep rates and the influence of different cooling procedures. Moreover, we present relaxation measurements $\kappa(t)$  performed after different field-sweep cycles and we discuss the influence of the direction of the heat current on the hysteresis effects for  \mbox{$\vec H \,||\, [111]$}.

\section{Experimental}

Large single crystals of $\text{Dy}{}_\text{2}\text{Ti}_\text{2}\text{O}_\text{7}$ were grown by the floating-zone technique in a mirror furnace. Oriented crystals of approximate dimensions \mbox{$3 \times 1 \times 1 \, \text{mm}^3$} were used to measure the thermal conductivity by the standard steady-state method. A heater produced a temperature gradient within the sample which was measured by two calibrated \mbox{$\text{RuO}_{\text{2}}$} thermometers. The heat current $\vec{\jmath}$ was driven along the longest sample dimension, while the magnetic field was applied either parallel or perpendicular to $\vec{\jmath}$. Note that the standard steady-state method is a step-by-step technique where every data point needs several minutes to stabilize the temperature gradient and the average temperature. This then typically results in effective field-sweep rates of \mbox{$\approx 0.01 \, \text{T/min}$}. Most of the measurements were performed using this step-by-step technique, but, in addition, we also performed some measurements where the field was continuously varied with a larger rate of \mbox{$0.1 \, \text{T/min}$}.

Demagnetization effects were taken into account for all measurements. Due to the geometry with $\vec{\jmath}$ along the longest sample dimension this correction is large (up to \mbox{$\sim 0.5 \, \text{T}$}) for a magnetic field perpendicular to $\vec{\jmath}$, whereas much smaller corrections (\mbox{$\lesssim 0.08 \, \text{T}$}) are present for a field parallel to $\vec{\jmath}$. The demagnetization field is calculated on the basis of experimental magnetization data which were measured with a home-built Faraday magnetometer on thin samples to minimize the demagnetization effects within the magnetization measurements.

\section{Results}

\subsection{Magnetic field parallel $[001]$ and heat current along $[1\bar{1}0]$}

\begin{figure}
\begin{center}
\includegraphics[width=16cm]{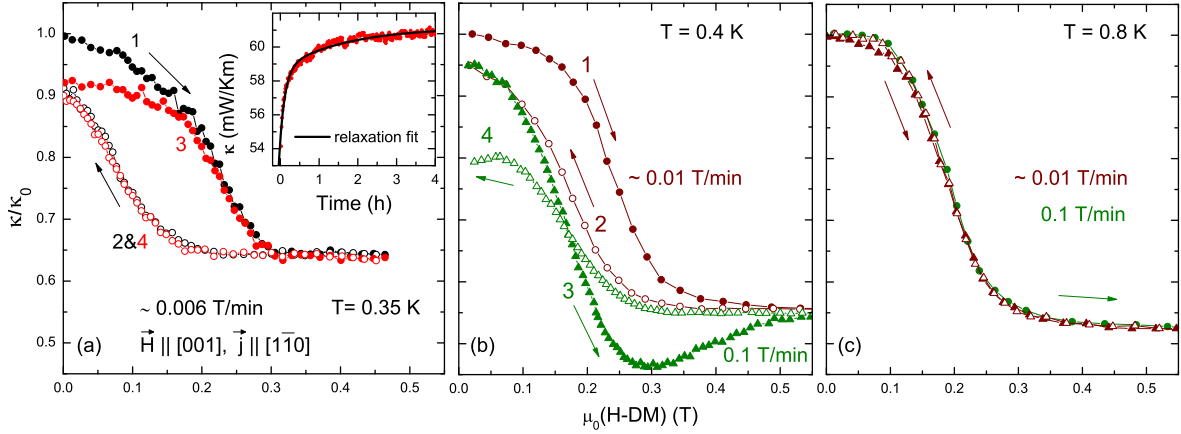}
\end{center}
\caption{(Color online) Field dependence \mbox{$\kappa(H)/\kappa_0$} for \mbox{$\vec H \,||\, [001]$} and \mbox{$\vec \jmath \,||\, [1\bar{1}0]$} for different field-sweep directions (marked by arrows) and field-sweep rates. The order of the successive field sweeps is marked by the numbers 1--4, where the initial sweep was started after cooling in zero field. The inset shows the relaxation of the reduced zero-field value towards the initial $\kappa_0$. The data of (a) are from Ref.~\cite{Kolland2012}.}
\label{relax_100_j110}
\end{figure}

Figures \ref{relax_100_j110}(a)-(c) display field dependent measurements $\kappa(H)$ for different magnetic-field-sweep rates normalized to $\kappa_0$ obtained by zero-field cooling. As already discussed in Ref. \cite{Kolland2012}, figure \ref{relax_100_j110}(a) reveals that after cycling the magnetic field up and down at \mbox{$0.35 \, \text{K}$} the final $\kappa(H\rightarrow 0)$ only recovers about 90\% of its initial zero-field $\kappa_0$, see curves (1) and (2). A subsequent field cycle results in \mbox{$\kappa(H)$} curves (3) and (4) with equal endpoints, where curves (2) and (4) perfectly match each other. As is shown in the inset of figure~\mbox{\ref{relax_100_j110}(a)}, the reduced zero-field $\kappa(H\rightarrow 0)$ slowly relaxes back to the zero-field-cooled $\kappa_0$. The fit yields large relaxation times \mbox{$\tau_1 \simeq 8\, \text{min}$} and \mbox{$\tau_2 \simeq 100 \, \text{min}$}. In contrast, no such slow relaxation effects are observed in the field range above \mbox{$0.3 \, \text{T}$}, where $\kappa(H)$ is non-hysteretic.

This slow relaxation in the hysteretic region raises the question, in how far the measured \mbox{$\kappa(H)$} curves also depend on the field-sweep rate. In Fig. \ref{relax_100_j110}(b), $\kappa(H)$ is shown for different field-sweep rates at \mbox{$T=0.4 \, \text{K}$}. After the initial field cycle with a sweep rate of \mbox{$\sim 0.01 \, \text{T/min}$} the $\kappa(H \rightarrow 0)$ value recovers about 95\% of the initial  $\kappa_0$. Increasing the field-sweep rate of the subsequent field cycle to \mbox{$0.1 \, \text{T/min}$} causes additional features in $\kappa(H)$. The $\kappa(H)$ curve~(3) measured with increasing field shows a minimum around  \mbox{$0.3 \, \text{T}$}, but nevertheless approaches the initial $\kappa(H)$ curve~(1) around \mbox{$0.55 \, \text{T}$}. Decreasing the field again results in the $\kappa(H)$ curve~(4), which is almost constant and close to curve~(2) until the field falls below 0.25~T, where an even larger hysteresis opens and after this faster field cycle $\kappa(H \rightarrow 0)$ only recovers about 80\% of the initial  $\kappa_0$.

The data of Figs. \ref{relax_100_j110}(a) and~(b) clearly show that even small field-sweep rates of \mbox{$\sim 0.01 \, \text{T/min}$} may be too large to reach equilibrium states in the low-temperature range of spin ice. As may be naturally expected,  this slow equilibration vanishes towards higher temperatures, as is shown in Fig.~\ref{relax_100_j110}(c). On the other hand, the hysteresis effects also disappear in the low-temperature range when the magnetic field is increased. This is not only suggested by the data of Figs.~\ref{relax_100_j110}(a) and~(b), but also follows from our additional data measured to higher fields, which are non-hysteretic and do not show such slow relaxation effects~\cite{Kolland2012,Kolland2013}. The fact that the relaxation/hysteresis effects rather rapidly vanish towards higher fields appears also natural, because the magnetization of spin ice is essentially saturated in this field range. One might suspect that the slow equilibration in the low-temperature/low-field region is a single-ion property of the large Ising spins of the Dy$^{3+}$ ions, which only slowly equilibrate  because of the rather large splitting to the higher-lying crystal-field states. This can be ruled out, however, from our measurements on the related half-doped material \mbox{$(\text{Dy}{}_\text{0.5}\text{Y}{}_\text{0.5})_\text{2}\text{Ti}_\text{2}\text{O}_\text{7}$}, which does not show such hysteresis effects~\cite{Kolland2013}. Thus, we conclude that the slow equilibration is a particular spin-ice feature.
 
\subsection{Magnetic field parallel $[111]$ and heat current along $[1\bar{1}0]$}

\begin{figure}
\begin{center}
\includegraphics[width=15.7cm]{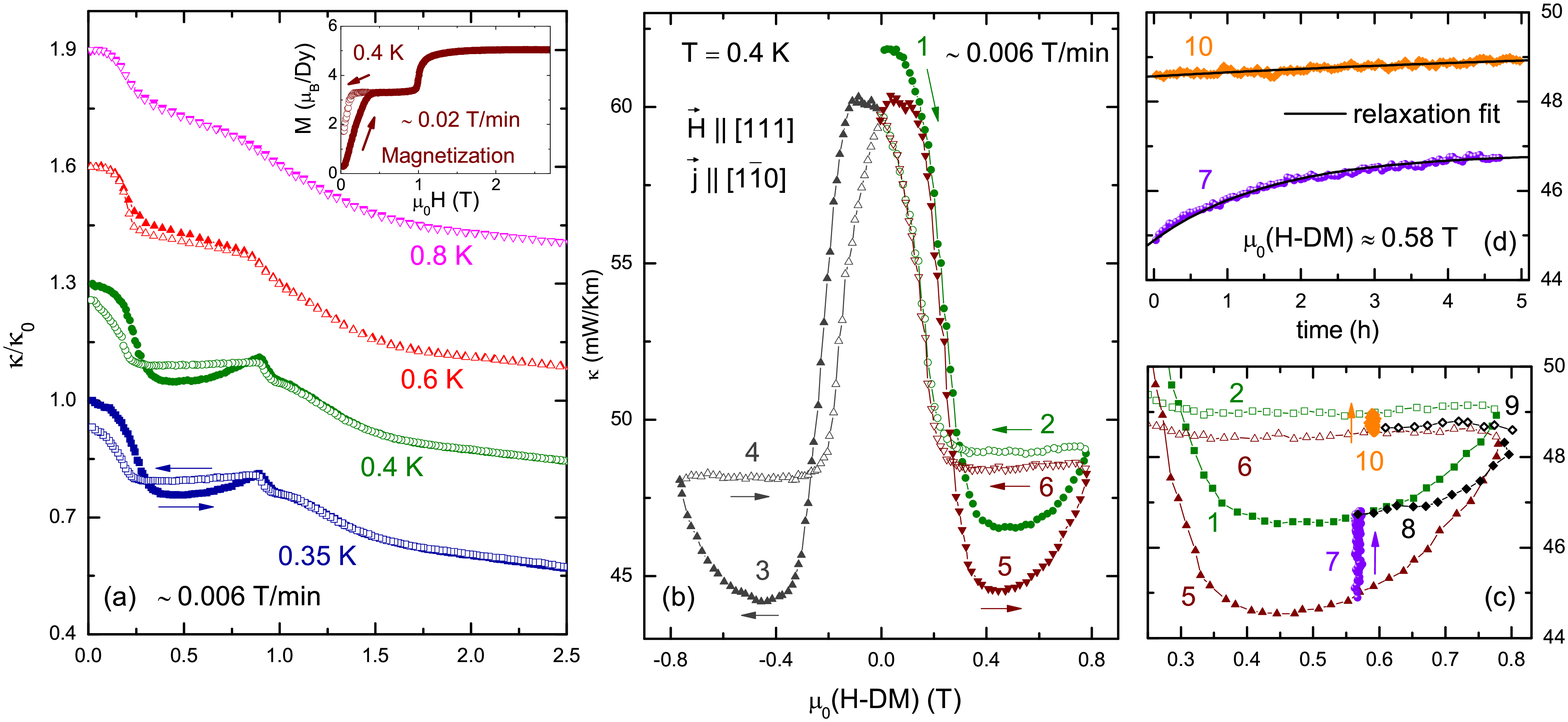}
\end{center}
\caption{(Color online) Field dependence $\kappa(H)$  for \mbox{$\vec H \,||\, [111]$} and \mbox{$\vec \jmath \,||\, [1\bar{1}0]$} for different field-sweep directions (marked by arrows). In (a), the \mbox{$\kappa(H)/\kappa_0$} curves for different temperatures  are shifted with respect to each other and the inset displays a corresponding magnetization curve (data from~\cite{Kolland2013}). Low-field hysteresis loops of $\kappa(H)$ are displayed in (b), where the order of the successive field sweeps is marked by the numbers 1--6. An expanded view of $\kappa(H)$  in the kagome-ice region is displayed in (c), which also shows the $\kappa(t)$ curves (7) and (10) and the additional field sweeps (8) and (9); see text. The relaxation curves $\kappa(t)$ with corresponding fits as a function of time are displayed in (d).}
\label{relaxation_B_111_j_110}
\end{figure}

Figure \ref{relaxation_B_111_j_110}(a) displays \mbox{$\kappa(H)/\kappa_0$} for \mbox{$\vec H \,||\, [111]$} with a heat current $\vec \jmath$ measured along $[1\bar10]$ at various temperatures. The kagome-ice phase is clearly seen in a plateau-like feature of $\kappa(H)$ below \mbox{$\sim 1 \, \text{T}$}. As already mentioned, below \mbox{$0.6 \, \text{K}$} the $\kappa(H)$ data show a clear hysteresis in the kagome-ice phase, whereas no such hysteresis is present in the plateau region of the corresponding magnetization curve as is shown in the inset of Fig.~\ref{relaxation_B_111_j_110}(a). Moreover, compared to the initial $\kappa_0$ obtained by zero-field cooling, the low-temperature field cycles result in reduced zero-field values $\kappa(H \rightarrow 0)$, which as a function of time slowly relax towards $\kappa_0$ (not shown). This zero-field relaxation is present for all three field directions \mbox{$\vec H \,||\, [001]$}, \mbox{$\vec H \,||\, [110]$},  and \mbox{$\vec H \,||\, [111]$}, where in all cases the heat current was driven along $[1\bar{1}0]$, see Ref.~\cite{Kolland2013}.

In order to further investigate the relaxation processes within the kagome-ice phase, complete low-field hysteresis loops of $\kappa(H)$ for positive and negative magnetic field were performed, which are shown in Fig.~\ref{relaxation_B_111_j_110}(b). Concerning the reduced zero-field values, the observed systematics of $\kappa(H)$ is analogous to that already discussed above for \mbox{$\vec H \,||\, [001]$}. In addition, we find that the hysteresis in the kagome-ice phase in the initial field cycle is less pronounced than in the subsequent cycles. In particular, the minima in the $\kappa(H)$ curves (3) and (5) are more pronounced than the minimum in the initial curve (1). Therefore, we also performed relaxation studies in this field region, which are shown in Figs.~\ref{relaxation_B_111_j_110}(c) and~(d). First, we measured the time dependence $\kappa(t)$ at constant field and temperature starting from the $\kappa(H)$ curve (5). Similar to the zero-field case, we observe a slow relaxation with a time constant \mbox{$\tau \approx 100\,\text{min}$} and the $\kappa(t)$ curve seems to relax towards the initial $\kappa(H)$ curve (1). Thus, the relaxation measurement was stopped after 4.5 hours and the field was cycled up to 0.8~T and back to $\sim 0.58 \, \text{T}$. The corresponding $\kappa(H)$ curves (8) and (9) do, however, not follow the initial field dependent $\kappa(H)$ curves (1) and (2). Instead, the $\kappa(H)$ curve (8) approaches curve (5) and the field-decreasing $\kappa(H)$ curve (9) essentially follows curve (6). At $\sim 0.58 \, \text{T}$, we then again measured $\kappa(t)$ and observed a weak, almost linear increase of $\kappa(t)$ yielding a very large relaxation time \mbox{$\tau \approx 520 \,\text{min}$}. These data suggest that the relaxation within the kagome-ice phase is not directly related to the zero-field relaxation and also reveal that, depending on the field and temperature region, a true thermal equilibrium can be hardly reached under typical experimental conditions.

\subsection{Magnetic field parallel $[111]$ and heat current along $[111]$}

\begin{figure}[t]
\begin{center}
\includegraphics[width=15.5cm]{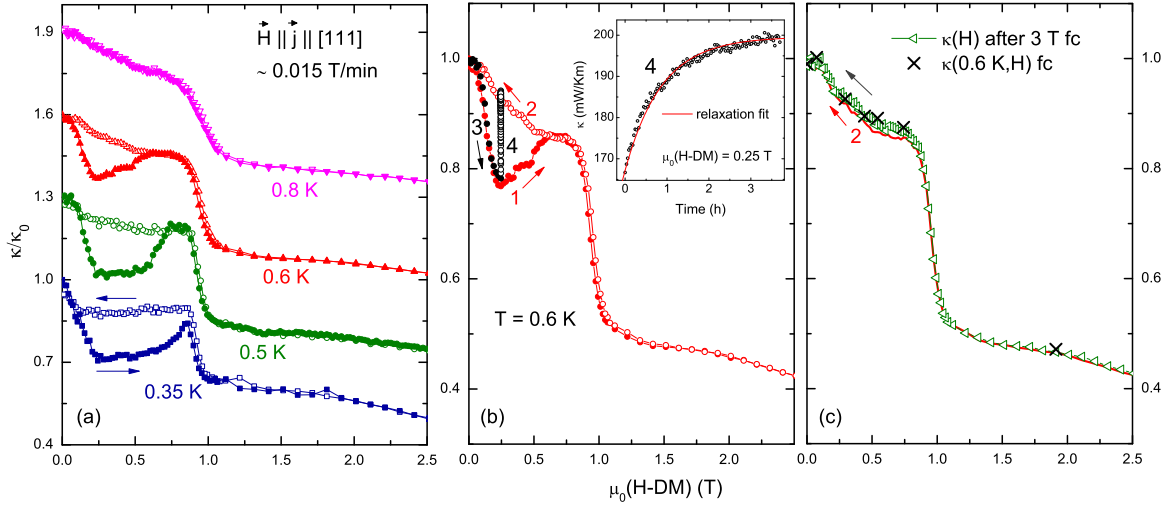}
\end{center}
\caption{(Color online) Field dependence \mbox{$\kappa(H)/\kappa_0$} for \mbox{$\vec H \,||\, [111]$} and \mbox{$\vec \jmath \,||\, [111]$} for different field-sweep directions (marked by arrows). The curves for different temperatures in (a) are shifted with respect to each other. Successive field sweeps (1) -- (3) at 0.6~K together with a relaxation curve (4) are shown in (b). The inset displays the $\kappa(t)$ curve (4) as a function of time,  while in (c) the \mbox{$\kappa(H)/\kappa_0$} curve (2) is compared to different \mbox{$\kappa(H)/\kappa_0$} values obtained by field cooling and to a field-decreasing run measured after field cooling in $3 \, \text{T}$.}
\label{kappa_B_j_111}
\end{figure}

Figure~\ref{kappa_B_j_111}(a) displays the field-dependence \mbox{$\kappa(H)/\kappa_0$} for the same magnetic-field direction $[111]$ as in the previous subsection, but now with the heat current \mbox{$\vec \jmath \,||\, [111]$}. Again, there is a hysteresis of $\kappa(H)$ within the kagome-ice phase which has no analogue in the magnetization curves, but there are also systematic differences between \mbox{$\kappa(H)/\kappa_0$} for the same field but different heat-current directions. First of all, the hysteresis width with respect to the different absolute values of $\kappa(H)$ is significantly larger for \mbox{$\vec \jmath \,||\, [111]$} than for \mbox{$\vec \jmath \,||\, [1\bar{1}0]$}. Secondly, the upper critical field, where the hysteresis closes, strongly decreases with increasing temperature for \mbox{$\vec \jmath \,||\, [111]$}, while it is essentially temperature-independent for \mbox{$\vec \jmath \,||\, [1\bar{1}0]$}. Finally, for \mbox{$\vec \jmath \,||\, [111]$} the intial zero-field $\kappa_0$ is recovered in the field-decreasing runs despite the fact that there is a finite remnant magnetization, when $M(H)$ is measured with a comparable sweep rate, see inset of Fig.~\ref{relaxation_B_111_j_110}(a).

Relaxation and zero-field-cooled measurements of $\kappa(H)$ reveal that in the hysteresis region of \mbox{$\kappa(H)/\kappa_0$} the upper branches, which are obtained with decreasing field, are closer to thermal equilibrium than the lower ones. For example, $\kappa(t)$ relaxes from the lower $\kappa(H)$ curve~(1) to the upper curve~(2) at $T=0.6$~K with a  relaxation time \mbox{$\tau \approx 52 \,\text{min}$}, as is shown in Fig.~\mbox{\ref{kappa_B_j_111}(b)}. Moreover, the $\kappa(H)$ values obtained either by field-cooling in various fixed fields or by field-cooling in 3~T and decreasing the field to zero well agree with the initial field decreasing $\kappa(H)$ curve~(2), see Fig.~\mbox{\ref{kappa_B_j_111}(c)}. Qualitatively, this hysteresis and relaxation of $\kappa(H)$ can be understood by assuming that field-induced disorder within the kagome-ice phase causes an additional suppression of $\kappa$, because starting from the fully polarized high-field state will cause less disorder in the kagome-ice phase than entering this phase from the entropic zero-field spin-ice ground state; see also the discussions in Refs.~\cite{Sun2013,Kolland2013}. Finally, our finding that the presence or absence of a zero-field hysteresis of $\kappa(H\rightarrow 0)$ for \mbox{$\vec H \,||\, [111]$} depends on the direction of $\vec \jmath $ indicates that the strength and the direction of a heat current through spin ice may be an additional parameter to influence its thermal equilibration.

\section{Conclusion}

The thermal conductivity $\kappa$ of the spin-ice compound $\text{Dy}{}_\text{2}\text{Ti}_\text{2}\text{O}_\text{7}$ shows strong hysteresis and slow relaxation processes towards equilibrium states in the low-temperature and low-field regime. In general, the thermal conductivity in the hysteretic regions slowly relaxes towards larger values suggesting that there is an additional suppression of the heat transport by field-induced disorder in the non-equilibrium states. The degree of hysteresis does not only depend on temperature and the magnetic-field direction, but also on the rate of the magnetic-field change and, for \mbox{$\vec H \,||\, [111]$} even on the direction of the heat current. The observation that the rate of thermal equilibration in spin ice can be influenced by a finite heat current along certain directions may possibly yield important information about the dynamics of monopole excitations and their interaction with phonons. However, further investigations of this effect for other directions of the magnetic field and different directions of the heat current are necessary.   
\newline
\newline
\textbf{Acknowledgments}
\newline
This work has been financially supported by the Deutsche Forschungsgemeinschaft via SFB 608 and the project LO 818/2-1.

\end{document}